\documentstyle[twocolumn,eqsecnum,aps]{revtex}
\begin{document}
\draft
\title{Fermions without vierbeins in curved
space-time}
\author{H. Arthur Weldon}
\address{Department of Physics, West Virginia University, Morgantown,
West Virginia
26506-6315}
\date{September 21, 2000}

\maketitle

\begin{abstract}
A general formulation of spinor fields in   Riemannian
space-time is  given without using vierbeins. The space-time
dependence of
the Dirac matrices required by the anticommutation relation
$\{\gamma_{\mu},\gamma_{\nu}\}=2g_{\mu\nu}$ determines the spin
connection. The action is invariant under any local spin base
transformations in the 32 parameter group Gl(4,c) and not  just
under local
Lorentz transformations. The  Dirac equation and the
 energy-momentum tensor are computed from the action.
\end{abstract}

\section{Introduction}

The Dirac equation for spinor fields was formulated in curved
space in 1929 by Fock and Ivanenko \cite{Fock}, who found it
convenient to employ vierbeins. That method provides a particular
  solution for
the space-time dependent Dirac matrices
$\gamma^{\nu}$  satisfying the anticommutation relation
\begin{equation}
\big\{\gamma^{\mu},\gamma^{\nu}\big\}
=2g^{\mu\nu}I.\label{anti}\end{equation}
 The
description of fermions in curved space-time using vierbeins
can be found in books by deWitt
\cite{B1}, Weinberg
\cite{B2}, Burell and Davies
\cite{B3}, and Wald \cite{B4}.
It frequently implied, and sometimes explicitly stated,
that fermionic fields cannot be formulated in curved space without
the use of vierbeins. This paper will show that
vierbeins are not necessary.

The reason for eschewing vierbeins is not mere novelty.
Always using vierbeins to
treat fermionic fields is rather like always using the Coulomb
gauge to perform electrodynamics calculations in Minkowski
space.  One can get the right answers in Coulomb
gauge. However it would be very puzzling if physics could
only be done in that gauge because the Coulomb gauge condition
 only holds in one Lorentz frame,
a requirement which is contrary to the principle
that all inertial frames are equivalent.

The vierbeins are the gravitational analogue of the Coulomb
gauge.
 At each point $x$ in a
curved space-time it is always possible to choose a set of four
locally inertial coordinates $\xi^{a}$, which depend on $x$.
The vierbeins are defined as the partial derivatives
\begin{equation}
e_{\mu}^{a}(x)={\partial\xi^{a}\over\partial x^{\mu}}.
\label{vb1}\end{equation}
Since coordinates that are inertial at one point will not be
inertial at a nearby point, the vierbiens have the property
that
$\partial_{\nu}e_{\mu}^{a}\neq\partial_{\mu}e_{\nu}^{a}$.
They  automatically satisfy the two relations
$e_{\mu}^{a}e_{\nu}^{b}\eta_{ab}=g_{\mu\nu}$ and
$e_{\mu}^{a}e_{\nu}^{b}g^{\mu\nu}=\eta^{ab}$.
The inverse of Eq. (\ref{vb1}) is
\begin{equation}
e^{\mu}_{a}(x)=g^{\mu\nu}e_{\nu}^{b}\,\eta_{ba}.
\label{vb2}\end{equation}
 From any set of space-time independent
Dirac matrices satisfying
\begin{displaymath}\big\{\gamma^{a},\gamma^{b}\big\}=2\eta^{ab}I,
\end{displaymath}
one can  solve  Eq. (\ref{anti}) by choosing the space-time
dependent Dirac matrices as particular linear combinations
of the constant Dirac matrices:
\begin{equation}
\gamma^{\mu}(x)=e^{\mu}_{a}(x)\gamma^{a}.
\label{14}\end{equation}
This is not the most general solution to Eq.
(\ref{anti}) but it does work. The action for the fermion fields
and the Dirac equation is expressed in terms of these vierbeins
\cite{B1,B2,B3,B4}.  The vierbein method is used for both the
classical Dirac field \cite{Ex1,Ex2,Ex3,Ex4} and for the
quantized Dirac field \cite{Q1,Q2}.

There are several unattractive features of the vierbein
formulation. (1) The 10 independent components of the metric tensor are
replaced by the 16 components of the vierbeins. (2) It is necessary to
introduce  a special inertial frame at each point
contrary to the basic principles that led Einstein to construct general
relativity.
Neither the inertial frames nor the constant Minkowski metric
 is  necessary
for spin 0 or spin 1 fields or for gravity itself.
(3) When the vierbein solution  for the Dirac matrices in
 Eq. (\ref{14}) is
inserted into the formula
\begin{equation}
\gamma_{5}=-i{\sqrt{-g}\over
4!}\epsilon_{\alpha\beta\mu\nu}
\gamma^{\alpha}\gamma^{\beta}\gamma^{\mu}\gamma^{\nu},
\label{gamma5}\end{equation}
all the space-time dependence cancels and  $\gamma_{5}$  becomes equal
to the constant matrix $\gamma_{(5)}$ from Minkowksi space.
Because of intrinsic parity violation, the matrix
$\gamma_{5}$ plays a central role in  the electroweak interactions.
The coupling of
quarks and  leptons  to the weak vector bosons W$^{\pm}$ and Z$^{0}$
is through a linear combination of vector currents
$\overline{\psi}\gamma^{\mu}\psi$ and axial vector currents
$\overline{\psi}\gamma^{\mu}\gamma_{5}\psi$.
 In the vierbein formalism, local Lorentz
transformations  change the space-time dependence of the $\gamma^{\mu}$,
but   $\gamma_{5}$ remains the same constant matrix as in Minkowski space.

In the standard vierbein formulation of fermion fields
their are two types of transformations: general coordinate
transformations and local Lorentz transformations. The Lagrangian
density must be a scalar under both types of transformations.

\paragraph*{General Coordinate Transformations.}
Under a general transformation $x^{\mu}\to
\widetilde{x}^{\mu}(x)$ of the coordinate system, each of the
four vierbeins transforms as a coordinate vector
\begin{equation}\widetilde{e}^{\mu}_{\;a}(\widetilde{x})
={\partial
\widetilde{x}^{\mu}\over
\partial x^{\nu}}\,e^{\mu}_{\;a}(x).
\end{equation}
The Dirac matrices transform as a vector:
\begin{mathletters}\begin{equation}
\widetilde{\gamma}^{\mu}(\widetilde{x})= {\partial
\widetilde{x}^{\mu}\over
\partial x^{\nu}}\;\gamma^{\nu}(x).
\end{equation}
The transformations matrices $\partial\widetilde{x}^{\mu}
/\partial x^{\nu}$ are 4$\times$4 real matrices belonging to the group
Gl(4,r). Because Gl(4,r) does not contain as a subgroup the spinor
representations of the Lorentz group \cite{Q1,Cartan}, it is
not possible for the spinor field $\psi(x)$ to transform
under general coordinate transformations. Therefore  the spinor
field transforms as a scalar under general coordinate
transformations:
\begin{equation}
\widetilde{\psi}(\widetilde{x})=\psi(x).\end{equation}
\end{mathletters}
This is completely standard \cite{B1,B2,B3,B4}.

\paragraph*{Local Lorentz Transformations.}  Conventionally the way
to introduce spin base transformations that act on the
components of the spinor quantities is to mimic Minkowski
space. Under local Lorentz transformations of the inertial
coordinate
$\xi^{a}\to\Lambda^{a}_{\;\;b}\xi^{b}$ the vierbeins mix
according to
\begin{displaymath}
e^{\prime\mu}_{a}(x)= e_{b}^{\mu}(x)\;\Lambda^{b}_{\;\;
a}.\end{displaymath}
The constant Dirac matrices
$\gamma^{a}$ transform  according to spinor representations of the
local Lorentz group:
\begin{displaymath}
\Lambda^{b}_{\;\;a}\gamma^{a}=S_{\rm Lor}(\Lambda)\gamma^{b} S_{\rm
Lor}(\Lambda).\end{displaymath}
Therefore under local Lorentz transformations of the inertial
coordinate, the Dirac matrices and spinor field transform as
\begin{mathletters}\begin{eqnarray}
\gamma^{\prime\mu}(x)=  && S_{\rm
Lor}(\Lambda)\gamma^{\mu}(x)S_{\rm Lor}^{-1}(\Lambda)\\
\psi^{\prime}(x)= && S_{\rm Lor}(\Lambda)\psi(x).
\end{eqnarray}\end{mathletters}
The vierbein does not appear explicitly in the
transformation equations. One may summarize this approach by saying
 that there
are two  types of transformations and the action must be a scalar under
both. Coordinate transformations employ real 4$\times$4 matrices
$\partial\widetilde{x}^{\mu}/\partial x^{\nu}$.
The spin base transformations
employ complex 4$\times$4 matrices $S_{\rm Lor}(\Lambda)$ that
 are spinor
representations of the  6-parameter Lorentz group.

Given the necessity of separate laws for general coordinate
transformations and for spin base transformations, it seems natural to
allow any complex 4$\times$4  matrix as a spin base transformation.
To do this  requires abandoning the vierbeins and dispensing with the
privileged position of local inertial frames and local Lorentz
transformations.

The remainder of the paper formulates the Dirac spinor field
in a Riemannian space without using vierbeins.
Section II develops two essential ingredients: the spin
metric $h$ which allows $\psi^{\dagger}h\psi$ to be invariant
under spin base transformations and the spin
connection $\Gamma_{\mu}$ which makes
$(\partial_{\mu}+\Gamma_{\mu})\psi$ transform covariantly
under a general spin base transformation \cite{Loos}.
 Section III treats the
equations of motion for the fermion field.  Requiring
the action to be
stationary under variations in
$\psi$ gives the Dirac equation. Requiring it to be stationary
 under variations
of $g^{\mu\nu}$ gives the Einstein field equation containing the
energy-momentum tensor of the fermions. Section IV provides a
summary. There
are five appendices. Appendix A summarizes the Clifford
 algebra basis.
Appendix B proves a general theorem which is employed in
 the development of the
spin connection, of the spin metric, and of the
energy-momentum tensor. Appendix C shows which components
 of the general spin
connection actually appear in the action for the fermion.
Appendix D develops the second order Dirac equation and relates
the spin curvature
tensor to the Riemann-Christoffel event curvature tensor.
 Appendix E shows that with
any fixed set of Dirac matrices satisfying Eq. (\ref{anti})
how to construct  a spin
base transformation to change the Dirac matrices to   a vierbein basis.

\section{Basic structure}

In the following it is not necessary to have an explicit
 form for the Dirac
matrices which satisfy the basic anticommutation relation,
Eq. (\ref{anti}).

\subsection{Spin metric}

The  matrices $\gamma^{\mu\dagger}$, which result from taking the
transpose and the complex conjugate of $\gamma^{\mu}$,
automatically  satisfy the same anticommutation relation.
As proven in Appendix B, the two solutions
must be related by a matrix transformation:
\begin{equation}
\gamma^{\mu\dagger}=h\,\gamma^{\mu}\,
h^{-1},
\label{21}\end{equation}
for some matrix $h$.
The adjoint of  this relation  can be used to obtain
\begin{displaymath}
\big[h^{-1}h^{\dagger},\gamma^{\mu}\big]=0
\end{displaymath}
A basis for the  15 nontrivial matrices of the Clifford algebra can be
formed from products of the $\gamma^{\mu}$. Since all of these
will commute with $h^{-1}h^{\dagger}$, the latter must be proportional
to the identity matrix:
$h^{-1}h^{\dagger}=zI$ for some complex number $z$. Taking the determinant
of both sides shows that $|z|=1$ and that $({\rm Det}\,h)^{2}=z^{-4}$. It
is convenient to redefine $h$ by changing
 $h\to h\sqrt{z}$. Then $h^{-1}h^{\dagger}=\sqrt{zz^{*}}=1$
so that  $h$ is hermitian:
\begin{equation}h^{\dagger}=h,
\end{equation}
and  ${\rm Det}\, h=1$.
The matrix $h$ will be the spin metric.
The definition in  Eq. (\ref{21}) also implies that
\begin{displaymath}
\gamma_{5}^{\dagger}=-h\gamma_{5}h^{-1}\end{displaymath}
It is helpful to note that the matrices
$h\gamma^{\mu}$ and
$h\gamma^{\mu}\gamma_{5}$ are  hermitian, whereas
$h\gamma_{5}$ and $h[\gamma^{\mu},\gamma^{\nu}]$ are
anti-hermitian.

\subsection{Spin connection}

Covariant derivatives of tensor quantities require the
 Riemann-Christoffel event
connection
\begin{equation}
\Gamma^{\nu}_{\mu\lambda}={1\over 2}g^{\nu\alpha}
\big(\partial_{\mu}g_{\alpha\lambda}+\partial_{\lambda}g_{\alpha\mu}
-\partial_{\alpha}g_{\mu\lambda}\big).\label{event}
\end{equation}
Covariant derivatives of spinors  require the spin connection.
Whatever choice is made for the Dirac matrices $\gamma^{\nu}$,
one can compute the derivatives
$\partial_{\mu}\gamma^{\nu}$. From these derivatives one can compute
the coefficients
\begin{mathletters}\begin{eqnarray}
t_{\mu}^{\alpha\beta}=&&{-1\over 32}{\rm Tr}\Big(\gamma^{\alpha}
\,(\partial_{\mu}\gamma^{\beta}+\Gamma_{\mu\lambda}^{\beta}
\gamma^{\lambda})\Big)\\
v_{\mu}^{\alpha}=&&{1\over 48}{\rm
Tr}\Big([\gamma^{\alpha},\gamma_{\nu}]\,\partial_{\mu}\gamma^{\nu}
\Big)\\
a_{\mu}^{\alpha}=&&{1\over 8}{\rm
Tr}\big(\gamma_{5}\partial_{\mu}\gamma^{\alpha}\big)\\
 p_{\mu}=&&{1\over
32}{\rm Tr}\big(\gamma_{5}\gamma_{\nu}\partial_{\mu}\gamma^{\nu}\big).
\end{eqnarray}\label{tvap}\end{mathletters}
It is easy to check the antisymmetry
$t_{\mu}^{\alpha\beta}+t_{\mu}^{\beta\alpha}=0$. For a fixed choice
of
$\mu$ there  are
 15 complex coefficients: 6 independent
values of
$t_{\mu}^{\alpha\beta}$, 4 values of $v_{\mu}^{\alpha}$,
4 values of $a_{\mu}^{\alpha}$, and one $p_{\mu}$.
Under general coordinate transformations each of these transforms as a
tensor as indicated by the indices. (For comparison, with the vierbein
solution for the  Dirac matrices  the coefficients  $v_{\mu}^{\alpha}$,
$a_{\mu}^{\alpha}$, $p_{\mu}$, and
$s_{\mu}$ all vanish; and the nonvanishing term $t_{\mu}^{\alpha\beta}$
has 6 real coefficients.)

 The spin connection is
 $\Gamma_{\mu}$
has the general decomposition in terms Dirac matrices as
\begin{equation}
\Gamma_{\mu}=t_{\mu}^{\alpha\beta}
\big[\gamma_{\alpha},\gamma_{\beta}\big]
+v_{\mu}^{\alpha}\gamma_{\alpha}
+a_{\mu}^{\alpha}\gamma_{\alpha}\gamma_{5}
+p_{\mu}\gamma_{5}+s_{\mu}I,\label{spincon}
\end{equation}
where $s_{\mu}$ is undetermined.
 The letters
$t,v,a,p,s$ denote tensor, vector, axial vector,
pseudoscalar, and scalar.
 Appendix B shows that the spin
connection given in Eq. (\ref{spincon}) satisfies
\begin{equation}
\partial_{\mu}\gamma^{\nu}+\Gamma^{\nu}_{\mu\lambda}
\gamma^{\lambda}+\big[\Gamma_{\mu},\gamma^{\nu}\big]
=0.\label{Sch}\end{equation}
It is elementary that if $\Gamma_{\mu}$ is postulated to satisfy
Eq. (\ref{Sch}) then the coefficients are as given in Eq. (\ref{tvap}).
Appendix B proves the converse, that Eq. (\ref{Sch}) is actually satisfied.
The result is non-trivial and it is worth doing some counting to
appreciate what has happened. For a fixed choice of $\mu$ and $\nu$
the derivative $\partial_{\mu}\gamma^{\nu}$ will generally be a linear
combination of 15 Dirac matrices. If  only $\mu$ is fixed
and $\nu$ runs over its four values then Eq. (\ref{Sch}) is a set of 60
linear equations. These 60 equations are solved by the 15 coefficients
displayed in Eq. (\ref{tvap}).

 Eq. (\ref{Sch}) also determines the derivative of products of
Dirac matrices. In particular, it gives
\begin{displaymath}
\partial_{\mu}\gamma_{5}+\big[\Gamma_{\mu},\gamma_{5}\big]=0.
\end{displaymath}

\subsection{Spinor fields}

The covariant derivative of the fermion field $\psi$ is
\begin{equation}
\nabla_{\mu}\psi=\partial_{\mu}\psi+\Gamma_{\mu}\psi.
\label{covpsi}\end{equation}
The fermion action is
$A_{f}=\int d^{4}x\,\sqrt{-g}\;{\cal L}_{f}$, where the Lagrangian
density is
\begin{equation}
{\cal L}_{f}={i\over 2}\psi^{\dagger}h\gamma^{\mu}\nabla_{\mu}\psi
-{i\over 2}(\nabla_{\mu}\psi)^{\dagger}h\gamma^{\mu}\psi
-m\,\psi^{\dagger}h\psi.\label{action1}
\end{equation}
To examine the role of the spin connection it is helpful to write out
${\cal L}$ in detail:
\begin{eqnarray}
{\cal L}=&&{i\over 2}\psi^{\dagger}h\gamma^{\mu}\partial_{\mu}\psi
-{i\over 2}(\partial_{\mu}\psi)^{\dagger}h\gamma^{\mu}\psi
-m\,\psi^{\dagger}h\psi\nonumber\\
+&&{i\over 2}\psi^{\dagger}\Big(h\gamma^{\mu}\Gamma_{\mu}
-\Gamma_{\mu}^{\dagger}h\gamma^{\mu}\Big)\psi.\nonumber
\end{eqnarray}
When the full spin connection is substituted in the action there
are numerous cancellations so that the 16 complex parameters are reduced
to 16 real parameters. (See Appendix C.)  Of these cancellations,
the simplest
occurs in the the part of
$\Gamma_{\mu}$ that is proportional to the identity matrix, viz.
$s_{\mu}I$. In particular  $({\rm Re}\,s_{\mu})I$
does not appear in the action. It is very convenient to subtract off
 $2({\rm
Re}\, s_{\mu})I$ from the spin connection and define
\begin{mathletters}\begin{equation}
\widehat{\Gamma}_{\mu}=\Gamma_{\mu}-{1\over 4}{\rm Re}\,[{\rm
Tr}\,(\Gamma_{\mu})]I.\label{gamhat}
\end{equation}
 The subtracted spin connection satisfies
\begin{equation}
{\rm Re}\big[\,{\rm Tr}\big(\widehat{\Gamma}_{\mu}\big)\big]=0.
\label{real}\end{equation}\end{mathletters}
It will be convenient to use $\widehat{\Gamma}_{\mu}$
in Sec. III for the discussion of the Dirac equation and the energy-momentum
tensor.

\subsection{Spin base transformations}

Starting from a set of Dirac matrices $\gamma^{\mu}$ satisfying the
anticommutation relation Eq. (\ref{anti}) one can change to a new set
by a spin base transformation
\begin{equation}
\gamma^{\prime\nu}=S\gamma^{\nu} S^{-1},
\end{equation}
where $S$ is an invertible, complex $4\times 4$ matrix with
arbitrary dependence on space-time. Thus $S$ belongs to the 32 parameter group
 Gl(4,c).  Under a
spin base transformation the
 derivative of the Dirac matrices transforms
inhomogeneously:
\begin{displaymath}
\partial_{\mu}\gamma^{\prime\nu}=S\Big(\partial_{\mu}
\gamma^{\nu}
+\big[S^{-1}\partial_{\mu}S,\gamma^{\nu}\big]
\Big)S^{-1}
\end{displaymath}
The coefficients in Eq. (\ref{tvap}) also change inhomogeneously
and this gives the transformed spin connection
\begin{equation}
\Gamma_{\mu}^{\prime}=S\Gamma_{\mu}S^{-1}-S^{-1}\partial_{\mu} S.
\label{newcon}\end{equation}
  The transformed Dirac
matrix and the transformed spin connection satisfy
\begin{equation}
\partial_{\mu}\gamma^{\prime\nu}+\Gamma^{\nu}_{\mu\lambda}
\gamma^{\prime\lambda}+\big[\Gamma_{\mu}^{\prime},
\gamma^{\prime\nu}\big]
=0.\end{equation}
The event connection $\Gamma_{\mu\lambda}^{\nu}$  does not
change under a spin base transformation.

Under a spin base transformation the spinor field behaves as
\begin{mathletters}\begin{equation}
\psi^{\,\prime}=S\psi.\end{equation}
The covariant derivative defined in Eq. (\ref{covpsi}) transforms
homogeneously:
\begin{equation}
\nabla^{\,\prime}_{\mu}\psi^{\,\prime}=S\,\nabla_{\mu}\psi.
\end{equation}\end{mathletters}

 The adjoints of
the new Dirac matrices
 satisfy
\begin{mathletters}\begin{equation}
\gamma^{\prime\mu\dagger}=h^{\,\prime}\gamma^{\prime\mu}
h^{\,\prime -1},
\end{equation}
where the transformed spin metric is
\begin{equation}
h^{\,\prime}=S^{\dagger -1}h\, S^{-1}.\end{equation}\end{mathletters}
There are several comparisons to be made. (1) The fact that the spin
metric changes under a spin base transformation is  analogous to
the fact that   the event metric $g_{\mu\nu}$ changes as a result of
coordinate transformations.
(2) By such a transformation it is always possible to make
$h^{\prime}$ a constant matrix with a convenient form, e.g.
  diagonal. However subsequent spin base
transformations will change $h^{\prime}$, unless one
arbitrarily  limits the allowed  spin base transformations.
(3) One can rewrite the above relation as
\begin{displaymath}
S^{\dagger}=hS^{-1}h^{\prime -1}.\end{displaymath}
This form emphasizes the relation of $S^{\dagger}$ to $S^{-1}$.
If one artificially limits the the allowed spin base transformations to
those which do not change the spin metric, then $h^{\prime}=h$ so that
$S^{\dagger}$ is matrix equivalent  to $S^{-1}$. Such a
choice would limit $S$ to be in the 16-parameter, unitary group U(2,2)
\cite{U22}.   In what follows this restriction will not be made: $h$ and
$h^{\prime}$ will not be constant and the spin base transformations will not be
required to keep the spin metric invariant.

 Associated with each spinor field $\psi$ is the Dirac
adjoint field
\begin{equation}
\overline{\psi}=\psi^{\dagger}h
\end{equation}
The product $\overline{\psi}\psi$ is manifestly
self-adjoint.  Under a spin base transformation the Dirac adjoint field
$\overline{\psi}$ changes into
\begin{equation}
\overline{\psi}^{\,\prime}=\psi^{\,\prime\dagger}h^{\,\prime}
=\psi^{\dagger}h S^{-1}=\overline{\psi}S^{-1}
\end{equation}
This provides various invariants under spin base transformations:
\begin{eqnarray}
\overline{\psi}^{\,\prime}\psi^{\,\prime}
=&&\overline{\psi}\psi\nonumber\\
\overline{\psi}^{\,\prime}\gamma^{\prime}_{\nu}\nabla^{\prime}_{\mu}
\psi^{\,\prime}=
&&\overline{\psi}\gamma_{\nu}\nabla_{\mu}\psi\nonumber\\
\overline{\psi}^{\,\prime}\gamma_{\nu}^{\prime}\psi^{\,\prime}=
&&\overline{\psi}\gamma_{\nu}\psi\nonumber
\end{eqnarray}
The first of these is the mass
term in the fermion action; the second is part of the
energy-momentum tensor; and the third is the vector current
for electromagnetism.

\subsection{Additional covariant derivatives}
 The covariant derivatives of
various additional quantities will be needed. Since $\overline{\psi}\psi$
is  a coordinate scalar and a spin base scalar,  the Leibnitz
product rule gives
$\partial_{\mu}(\overline{\psi}\psi)
=(\nabla_{\mu}\overline{\psi})\psi+\overline{\psi}
(\nabla_{\mu}\psi)$.
Consequently the  covariant derivative of the Dirac
adjoint field is
\begin{equation}
\nabla_{\mu}\overline{\psi}=\partial_{\mu}\overline{\psi}
-\overline{\psi}\Gamma_{\mu}.\label{b}
\end{equation}
On the other hand, the adjoint of  Eq.
(\ref{covpsi}) is
\begin{equation}
\nabla_{\mu}\psi^{\dagger}=\partial_{\mu}\psi^{\dagger}+\psi^{\dagger}
\Gamma_{\mu}^{\dagger}\label{a}
\end{equation}
where $\Gamma_{\mu}^{\dagger}$ is the  complex
conjugate, transpose matrix. Comparing these last two
results gives the covariant derivative of the spin metric:
\begin{equation}
\nabla_{\mu}h=\partial_{\mu}h-h\Gamma_{\mu}-
\Gamma_{\mu}^{\dagger}h.
\end{equation}
It is possible to evaluate this.   Taking the complex
conjugate, transpose of Eq. (\ref{Sch}) and using the
definition of the spin metric  leads to
\begin{displaymath}
0=\Big[h^{-1}\partial_{\mu}h-\Gamma_{\mu}
-h^{-1}\Gamma^{\dagger}_{\mu}h,\,\gamma^{\nu}
\Big].\end{displaymath}
By Schur's lemma the only matrix than commutes with all the
$\gamma^{\nu}$ is a multiple of the identity matrix
so that
\begin{displaymath}
h^{-1}\partial_{\mu}h-\Gamma_{\mu}
-h^{-1}\Gamma^{\dagger}_{\mu}h=c_{\mu}I
\end{displaymath}
One can evaluate $c_{\mu}$  by taking the the trace of both
sides of this relation:
\begin{displaymath}
c_{\mu}=-{1\over 2}{\rm Re}\,[{\rm  Tr}(\Gamma_{\mu})].
\end{displaymath}
In terms of  the subtracted spin connection
$\widehat{\Gamma}_{\mu}$ defined in Eq. (\ref{gamhat}) this reads
\begin{mathletters}\begin{equation}
h^{-1}\partial_{\mu}h-\widehat{\Gamma}_{\mu}
-h^{-1}\widehat{\Gamma}^{\dagger}_{\mu}h=0.\label{covh}
\end{equation}
This relation is also useful as a way of computing the adjoint of the
spin connection:
\begin{equation}
\widehat{\Gamma}_{\mu}^{\dagger}=-h\widehat{\Gamma}_{\mu}h^{-1}
+(\partial_{\mu}h)h^{-1}.
\end{equation}
A third possibility is to view Eq. (\ref{covh}) as the vanishing of the
covariant derivative of the spin metric:
 \begin{equation}
\widehat{\nabla}_{\mu}h=\partial_{\mu}h-h\widehat{\Gamma}_{\mu}
-\widehat{\Gamma}_{\mu}^{\dagger}h=0.
\end{equation}\end{mathletters}

\section{Field equations}

The action for the fermion field is
$A_{f}=\int d^{4}x\,\sqrt{-g}\;{\cal L}_{f}$, where the Lagrangian
density for fermions  is
\begin{equation}
{\cal L}_{f}={i\over 2}\psi^{\dagger}h\gamma^{\mu}\nabla_{\mu}\psi
-{i\over 2}(\nabla_{\mu}\psi)^{\dagger}h\gamma^{\mu}\psi
-m\overline{\psi}\psi.\label{action2}
\end{equation}
The independent variables are the fermion field $\psi$,
its adjoint $\psi^{\dagger}$
and the  metric tensor $g^{\mu\nu}$. The Dirac matrices,
the spin connection,
and the spin metric depend  on the event metric $g_{\mu\nu}$. There is
 no vierbein.  The Lagrangian density is a scalar
under general coordinate transformations and invariant under
general spin base transformations in Gl(4,c).

\subsection{Dirac equation}

As discussed in Sec. II C,
the real part of the trace of the spin connection $\Gamma_{\mu}$
does not contribute to the action and it is convenient to discard
that part by
employing $\widehat{\Gamma}_{\mu}$ as defined in Eq. (\ref{gamhat})
and define the matrix differential operator
\begin{equation}
K=ih\gamma^{\mu}(\partial_{\mu}+\widehat{\Gamma}_{\mu})-mh.\label{K}
\end{equation}
The fermion action can be
written
\begin{equation}
A_{f}=\int\! d^{4}x\sqrt{-g}\Big\{{1\over 2}
\psi^{\dagger}K\psi
+{1\over
2}(K\psi)^{\dagger}\psi\Big\},\label{action3}\end{equation}
By using the identity
\begin{displaymath}
\partial_{\mu}\big(\sqrt{-g}\,h\gamma^{\mu}\big)
=\sqrt{-g}\big(h\gamma^{\mu}\widehat{\Gamma}_{\mu}+
\widehat{\Gamma}_{\mu}^{\dagger}h\gamma^{\mu}\big),
\end{displaymath}
it is easy to show that for any two spinor fields $\psi$ and
$\chi$ that fall-off sufficiently fast ,
\begin{equation}
\int d^{4}x\sqrt{-g}\;\chi^{\dagger}K\psi
=\int d^{4}x \sqrt{-g}\;(K\chi)^{\dagger}\psi.
\label{selfadjoint}\end{equation}
Thus
$\sqrt{-g}\,K$ is a self-adjoint operator.

Extremizing  the action with respect to $\psi^{\dagger}$ gives
the generalized Dirac equation $K\psi=0$, or more explicitly
\begin{mathletters}\begin{equation}
i\gamma^{\mu}(\partial_{\mu}+\widehat{\Gamma}_{\mu})\psi=m\psi.
\label{Direqn}\end{equation}
Varying the action with respect to $\psi$ gives
\begin{equation}
-i\big(\partial_{\mu}\overline{\psi}\big)\gamma^{\mu}
+i\overline{\psi}\widehat{\Gamma}_{\mu}\gamma^{\mu}=m\overline{\psi}.
\end{equation}\end{mathletters}
The second equation is also implied by the first.
As a consequence of these the vector current
and the axial vector current obey the following:
\begin{eqnarray}
&&\partial_{\mu}(\overline{\psi}\gamma^{\mu}\psi)
+\Gamma_{\mu\lambda}^{\mu}
\overline{\psi}\gamma^{\lambda}\psi=0\nonumber\\
&&\partial_{\mu}(\overline{\psi}\gamma^{\mu}\gamma_{5}\psi)
+\Gamma_{\mu\lambda}^{\mu}
\overline{\psi}\gamma^{\lambda}\gamma_{5}\psi=i2m\,\overline{\psi}\psi,
\nonumber\end{eqnarray}
with the axial anomaly omitted.
Appendix D iterates the Dirac equation (\ref{Direqn}) to obtain
 the second order
form. The anticommutator of the covariant derivatives
$[\nabla_{\mu},\nabla_{\nu}]\psi$ introduces the spin curvature
 tensor, which is
directly related to the Riemann-Christoffel curvature tensor, so
 as to simplify the
second order wave equation.

\subsection{Energy-momentum tensor}

The energy-momentum tensor, being the source of the
gravitational field, must also be computed.
The fermion contribution to the energy-momentum tensor
requires varying the fermion action with respect to general
changes in the metric tensor:
\begin{displaymath}
\delta S={1\over 2}\int d^{4}x\sqrt{-g}\;(\delta g^{\mu\nu})
T_{\mu\nu}.\end{displaymath}

1. The theorem  proven in
Appendix B  provides the variational derivative of the
Dirac matrices, of the spin metric, and of the spin
connection. The most general  change in the
Dirac matrices that can result from a change in
the metric tensor is
\begin{mathletters}\begin{equation}
\delta\gamma^{\mu}={1\over 2}(\delta g^{\mu\nu})\gamma_{\nu}
+\big[\gamma^{\mu},G\big],\label{dgamma}
\end{equation}
where $G$ is some $4\times 4$ matrix. Without loss of
generality one may restrict $G$ to be traceless since the
identity matrix commutes with $\gamma^{\mu}$.   The dependence of
the matrix
$G$ on the metric tensor will depend upon how the basic
anticommutator in Eq. (\ref{anti}) is solved. (Even for the vierbein solution
$G$ is not zero.)

2. To compute the variation in the spin metric
$h$, take the variation of Eq. (\ref{21}) and substitute Eq. (\ref{dgamma})
to get
\begin{displaymath}
0=\Big[h^{-1}\delta h-G-h^{-1}G^{\dagger}h,\gamma^{\mu}\Big].
\end{displaymath}
By Schur's lemma the only matrix that commutes with all the
$\gamma^{\mu}$ is a multiple of the identity.
Since ${\rm Tr}(h^{-1}\delta h)=0$ and ${\rm Tr}(G)=0$, the
quantity that commutes is traceless, which implies
\begin{equation}
\delta h=hG+G^{\dagger}h.\label{dh}
\end{equation}
The product
$h\gamma^{\mu}$ appears throughout the action. Its variation is
therefore
\begin{displaymath}
\delta(h\gamma^{\mu})={1\over 2}(\delta g^{\mu\nu})
h\gamma_{\nu}+h\gamma^{\mu}G+G^{\dagger}h\gamma^{\mu}.
\end{displaymath}

3. Next one needs the dependence of the spin connection
$\Gamma_{\mu}$ on the metric.
Varying Eq. (\ref{Sch}) with respect to the metric tensor gives
\begin{displaymath}
0=\nabla_{\mu}(\delta\gamma^{\nu})
+(\delta\Gamma^{\nu}_{\mu\lambda})
\,\gamma^{\lambda}
+\big[\delta\Gamma_{\mu},\gamma^{\nu}\big].
\end{displaymath}
The first term can be evaluated using Eq. (\ref{dgamma}) and
the fact that
$\nabla_{\mu}\!\gamma^{\lambda}=0$:
\begin{displaymath}
\nabla_{\mu}(\delta\gamma^{\nu})=
{1\over
2}(\nabla_{\mu}\delta g^{\nu\lambda})
\gamma_{\lambda}-\big[\nabla_{\mu}G,
\gamma^{\nu}\big].
\end{displaymath}
By taking the variation of
$0=\nabla_{\mu}\,g^{\nu\lambda}$
this can be expressed more explicitly as
\begin{displaymath}
\nabla_{\mu}\!(\delta\gamma^{\nu})=
-{1\over 2}
(\delta\Gamma^{\nu}_{\mu\lambda})\gamma^{\lambda}
-{1\over
2}(\delta\Gamma_{\mu\alpha}^{\lambda})g^{\alpha\nu}\gamma_{\lambda}
-\big[\nabla_{\mu}G,\gamma^{\nu}\big].
\end{displaymath}
Substituting  above gives
\begin{displaymath}
0={1\over 2}(\delta\Gamma^{\nu}_{\mu\lambda})\gamma^{\lambda}
-{1\over
2}(\delta\Gamma_{\mu\alpha}^{\lambda})g^{\alpha\nu}\gamma_{\lambda}
+\big[\Gamma_{\mu}\!-\!\nabla_{\mu}G,\gamma^{\nu}\big].
\end{displaymath}
The first two terms together can be written as a commutator with
$\gamma^{\nu}$ so that
\begin{displaymath}
0=\Big[\Gamma_{\mu}\!-\!\nabla_{\mu}G
-{1\over
8}\delta\Gamma_{\mu\beta}^{\alpha}\,\big[\gamma_{\alpha},
\gamma^{\beta}\big],\gamma^{\nu}\Big].
\end{displaymath}
Since each term on the left hand side of the commutator is
traceless, Schur's lemma implies that
\begin{equation}
\delta\Gamma_{\mu}=\partial_{\mu}G+
\big[\Gamma_{\mu},
G\big]+{1\over
8}(\delta\Gamma_{\mu\beta}^{\alpha})
\,\big[\gamma_{\alpha},\gamma^{\beta}\big].
\label{dGamma}
\end{equation}\end{mathletters}

4. To compute the energy-momentum tensor for a fermion
field requires varying the action given in Eq.
(\ref{action2}):
\begin{eqnarray}
\delta S=&&\int d^{4}x\;(\delta\sqrt{-g})\;
\Big\{{1\over 2}\psi^{\dagger}K\psi
+{1\over 2}(K\psi)^{\dagger}\psi\Big\}\nonumber\\
+&&\int d^{4}x\;\sqrt{-g}\,
\Big\{{1\over 2}\psi^{\dagger}(\delta K)\psi
+{1\over 2}\big((\delta
K)\psi\big)^{\dagger}\psi\Big\}\nonumber
\end{eqnarray}
The variation of the differential operator  $K$ given in
Eq. (\ref{K}) with respect to the metric tensor can be computed using
 Eqs. (\ref{dgamma}),
(\ref{dh}),  and(\ref{dGamma}) with the result:
\begin{eqnarray}
\delta K=&&{i\over 2}(\delta
g^{\mu\nu})h\gamma_{\nu}(\partial_{\mu}+
\widehat{\Gamma}_{\mu})
+KG+G^{\dagger}K\nonumber\\
&&+{i\over
8}(\delta\,\Gamma_{\mu\beta}^{\alpha})
\,h\gamma^{\mu}\big[\gamma_{\alpha},\gamma^{\beta}\big]
\label{dK}\end{eqnarray}
The Dirac equation, $K\psi=0$, and the self-adjoint property
Eq. (\ref{selfadjoint}) guarantees that the terms $KG$ and
$G^{\dagger}K$ will make no contribution. Consequently
it was never necessary to know the matrix $G$ explicitly.
The last term in Eq. (\ref{dK}) is anti-hermitian
and automatically cancels in $\delta S$.
Thus the variation gives
\begin{eqnarray}
\delta S=\!\int \!d^{4}x\sqrt{-g}\;\delta g^{\mu\nu}
\Big\{&&{i\over 4}\psi^{\dagger}h\gamma_{\nu}
\nabla_{\mu}\psi
-{i\over 4}(\nabla_{\mu}\psi)^{\dagger}
h\gamma_{\nu}\psi\Big\}
\nonumber\end{eqnarray}
Symmetrizing with respect to $\mu\nu$ gives the
final result for the energy-momentum tensor for any fermion
field :
\begin{eqnarray}
T_{\mu\nu}=&&{i\over 4}\big(\overline{\psi}\gamma_{\nu}
\nabla_{\mu}\psi+\overline{\psi}\gamma_{\mu}\nabla_{\nu}\psi\big)\\
-&&{i\over
4}\big((\nabla_{\mu}\overline{\psi})
\gamma_{\nu}\psi+(\nabla_{\nu}\overline{\psi})
\gamma_{\mu}\psi\big).
\end{eqnarray}
The Einstein field equations are, as always,
\begin{equation}
R_{\mu\nu}-{1\over 2}g_{\mu\nu}R=8\pi G\, T_{\mu\nu}.
\end{equation}

\section{Summary and Conclusions}

The action for fermions that has been constructed  is invariant
under two separate transformations: general coordinate
transformations and local spin base transformations.

\paragraph*{1. General Coordinate Transformations.}
Under a general transformation $x^{\mu}\to\widetilde{x}^{\mu}(x)$ of
the coordinate system the Dirac matrices and the fermion field
transform as
\begin{mathletters}\begin{eqnarray}
\widetilde{\gamma}^{\mu}(\widetilde{x})= &&{\partial
\widetilde{x}^{\mu}\over
\partial x^{\nu}}\;\gamma^{\nu}(x)\\
\widetilde{\psi}(\widetilde{x})= &&\psi(x).
\end{eqnarray}\end{mathletters}
This is completely standard.

\paragraph*{2. Spin Base Transformations.}
Under arbitrary spin base transformations
 the Dirac matrices and spinor field transform as
\begin{mathletters}\begin{eqnarray}
\gamma^{\prime\mu}(x)=  && S(x)\gamma^{\mu}(x)S^{-1}(x)\\
\psi^{\prime}(x)= && S(x)\psi(x),
\end{eqnarray}\end{mathletters}
where $S(x)$ is a any matrix in $Gl(4,c)$.

\paragraph*{3. Spin base transformations to vierbein base.}
According to the theorem proven in Appendix B, any two sets
of Dirac matrices that solve the anticommutation relations
 must be related by a spin-base similarity
transformation.  Consequently every solution to the
anticommuation relations is spin-base equivalent to
a vierbein solution.
Appendix E shows how, starting from an arbitrary set of Dirac matrices
$\gamma^{\mu}$, to construct a transformation matrix $S$ and the quantities
$E^{\mu}_{a}$ satisfying
\begin{equation}
\gamma^{\mu}=E^{\mu}_{a}\,S\gamma^{a}S^{-1}.
\end{equation}
The construction  does not prove that $E_{a}^{\mu}$ are
derivatives of a locally inertial coordinate as are the vierbeins in Eq.
(\ref{vb1}).

\paragraph*{4. Inclusion of gauge fields.} Every species of fermion field
contains the covariant derivative
$(\partial_{\mu}+\widehat{\Gamma}_{\mu})\psi$. Under a spin base
 transformation
every fermion field undergoes the same transformation $\psi\to S\psi$.
By contrast, nonabelian gauge transformations always transform
 each fermion in
a muliplet differently. For example,  in QCD each type of
quark transforms as a triplet of the SU(3) color group. In the electroweak
interactions the chiral projections ${1\over 2}(1-\gamma_{5})\psi$ of the
quarks and leptons transform as doublets of the gauge group SU(2) and these
gauge transformations are  distinct from spin base transformations.
However the full electroweak gauge group is SU(2)$\times$U(1) and
the abelian
U(1) factor requires some discussion.

It is simplest consider  QED and then return to the electroweak
U(1) gauge invariance.
Let $\psi_{e}$ be the field of the electron field; $\psi_{u}$, the field
of the up quark; and
$\psi_{d}$, the field of  the down quark.
 The kinetic terms in the action are
\begin{eqnarray}
\overline{\psi}_{e}(\partial_{\mu}+
\widehat{\Gamma}_{\mu}-ieA_{\mu})\psi_{e}
+&&\overline{\psi}_{u}(\partial_{\mu}+\widehat{\Gamma}_{\mu}+i{2\over
3}eA_{\mu})\psi_{u}\nonumber\\
+&&\overline{\psi}_{d}(\partial_{\mu}+\widehat{\Gamma}_{\mu}-i{1\over
3}eA_{\mu})\psi_{d}.\nonumber
\end{eqnarray}
Included among the spin base transformations is  the  phase change
$\psi\to\exp(-i\phi )\psi$ of all three fields,  where $\phi$
is an arbitrary
real function.
Under this spin base transformation $\widehat{\Gamma}_{\mu}\to
\widehat{\Gamma}_{\mu}+i(\partial_{\mu}\phi) I$ and $A_{\mu}$
 does not change.
By contrast, under an electromagnetic gauge
transformation each fermion field
transforms with a different phase: $\psi_{e}\to\exp(-i\phi )\psi_{e}$,
$\psi_{u}\to\exp(i{2\over 3}\phi )\psi_{u}$, and
 $\psi_{d}\to\exp(-i{1\over
3}\phi )\psi_{d}$. The vector potential changes, $A_{\mu}\to
A_{\mu}-\partial_{\mu} \phi/e$, but
$\widehat{\Gamma}_{\mu}$ does not change.

The behavior of the abelian group in the SU(2)$\times$U(1) electroweak
interactions is analogous except that there is parity violation in the
coupling to the U(1) vector potential $B_{\mu}$.
Included among the spin base
transformations
$\psi\to S\psi$ are those for which
$S^{-1}\partial_{\mu}S=i\partial_{\mu}\phi(aI+b\gamma_{5})$ with $a$ and
$b$ the same for all fermions.   The U(1) gauge transformation
 is distinct from
these spin base transformations in that the values of
$a$ and $b$ are different for each field type and consequently the vector
potential
$B_{\mu}$ changes but the spin connection does not.

\paragraph*{5. Non-Riemannian spaces.} This paper
treats only Riemann spaces.  In a non-Riemannian space there
 additional degrees
of freedom beyond the metric that determine the geometry.
 The full event
connection is the sum of the Christoffel connection
 and an additional event
connection. The full spin connection is the sum of the
 Riemann spin connection
used here and an additional piece representing the new
 degrees of freedom.  The
are no obstacles encountered in  extending the above
treatment to this more
general space without using vierbeins. The full Gl(4,c) invariance is
maintained.

\acknowledgments

It is a pleasure to thank Richard Treat for many instructive
discussions regarding
this paper. This work  was supported in part by National Science Foundation
grant PHY-9900609.

\appendix

\section{Clifford Algebra}

In the anticommutation relations Eq. (\ref{anti}) the off-diagonal
components of the general metric $g^{\mu\nu}$ mean that the
anticommutator is
never zero. In proving various results it is often much easier
 to deal with
one covariant index and one contravariant index so that
\begin{displaymath}
\big\{\gamma_{\mu},\gamma^{\nu}\big\}=2\delta_{\mu}^{\nu}I.
\end{displaymath}
Using this one can show that the space-time dependent matrix
$\gamma_{5}$ defined in Eq. (\ref{gamma5}) has the property
\begin{equation}
\gamma_{5}\gamma^{\nu}\gamma_{5}=-\gamma^{\nu}.\label{Q1}
\end{equation}
Contracting both sides with $\gamma_{\nu}$ gives
\begin{equation}
\gamma_{5}\gamma_{5}=I.\end{equation}
Taking  trace of Eq.(\ref{Q1}) and using  the cyclic property
 gives ${\rm Tr}(\gamma^{\nu})=-{\rm Tr}(\gamma^{\nu})$ and therefore
\begin{equation}
{\rm Tr}\,(\gamma^{\nu})=0.\label{Tr}
\end{equation}
Because $\gamma_{\nu}\gamma^{\nu}=4$, Eq. (\ref{Q1}) implies
$\gamma_{5}=-\gamma_{\nu}\gamma_{5}\gamma^{\nu}/4$.
Taking the trace of this and using the cyclic property gives
${\rm Tr} (\gamma_{5})=-{\rm Tr}(\gamma_{5})$ and therefore
\begin{equation}
{\rm Tr}\,(\gamma_{5})=0.\end{equation}

The four Dirac matrices are the bases for the Clifford algebra.
The vector space of this algebra is spanned by 16 matrices which can be
chosen as the covariant tensors $1,\gamma_{\alpha},
[\gamma_{\alpha},\gamma_{\beta}], \gamma_{5}$, and
$\gamma_{\alpha}\gamma_{5}$. All except the identity are traceless.

\section{General theorem}

There are several computations which require knowing how the Dirac
matrices change when the metric tensor changes in a specified way.
The answer to this comes from the fundamental anticommutation
relation
\begin{equation}
\big\{\gamma^{\nu},\gamma^{\kappa}\big\}=2g^{\nu\kappa}I.
\label{Banti}\end{equation}
Under an infinitesimal change in the metric tensor  this
becomes
\begin{equation}
\big\{\Delta\gamma^{\nu},\gamma^{\kappa}\big\}
+\big\{\gamma^{\nu},\Delta\gamma^{\kappa}\big\}=2\Delta
g^{\nu\kappa}\,I.
\label{BDelta}\end{equation}
This Appendix  proves that the most general solution for
$\Delta
\gamma^{\nu}$ is
\begin{equation}\Delta\gamma^{\nu}={1\over 2}\big(\Delta
g^{\nu\lambda}\big)\gamma_{\lambda}
-\big[M,\gamma^{\nu}\big],\label{Btheorem}\end{equation}
where $M$ is a $4\times 4$ matrix. The specific values of the
change  $\Delta g^{\nu\lambda}$ will determine the specific
value of $M$. Before proving the theorem it may be helpful to
see the actual uses of this theorem.

\subsection{The spin connection}

One application of the theorem is to compute the
space-time derivative of the Dirac matrices in terms of the
derivatives of the metric. In other words, express
$\Delta \gamma^{\nu}=dx^{\mu}
\partial_{\mu}\gamma^{\nu}$
in terms of
$\Delta
g^{\nu\lambda}=dx^{\mu}\partial_{\mu}g^{\nu\lambda}$.
In this case the matrix $M$ must also be proportional to the
coordinate differentials:
$M=dx^{\mu}M_{\mu}$. Since $\nabla_{\mu}g^{\nu\lambda}=0$
it follows that
\begin{displaymath}
\Delta g^{\nu\lambda}
=-dx^{\mu}\big(\Gamma_{\mu\kappa}^{\nu}g^{\kappa\lambda}
+\Gamma_{\mu\kappa}^{\lambda}g^{\kappa\nu}\big).
\end{displaymath}
Then Eq. (\ref{Btheorem}) can be rearranged as
\begin{eqnarray}
dx^{\mu}\big(\partial_{\mu}\gamma^{\nu}
+\Gamma_{\mu\lambda}^{\nu}\gamma^{\lambda}\big)
={1\over 2}&&dx^{\mu}
\big(\Gamma_{\mu\kappa}^{\nu}g^{\kappa\lambda}
-\Gamma_{\mu\kappa}^{\lambda}g^{\kappa\nu}\big)\gamma_{\lambda}
\nonumber\\
- && \,dx^{\mu}\big[M_{\mu},\gamma^{\nu}\big].\nonumber
\end{eqnarray}
Because the two terms on the right involving the event
connection are antisymmetric under
$\nu\leftrightarrow\lambda$, the entire right hand side can
be written as commutator with $\gamma^{\nu}$:
\begin{equation}
\partial_{\mu}\gamma^{\nu}
+\Gamma_{\mu\lambda}^{\nu}\gamma^{\lambda}
=-\big[\Gamma_{\mu},\gamma^{\nu}\big]
\end{equation}
The matrix $\Gamma_{\mu}$ is the spin connection:
\begin{equation}
\Gamma_{\mu}
={1\over 8}\Gamma_{\mu\beta}^{\alpha}\big[\gamma_{\alpha},
\gamma^{\beta}\big]+M_{\mu}.
\end{equation}
Although $M_{\mu}$ is not a vector under general coordinate
transformations, the spin connection $\Gamma_{\mu}$
is automatically a vector.

\subsection{Pauli's theorem in curved space}

Given one set of Dirac matrices $\gamma^{\nu}$ satisfying
Eq. (\ref{Banti}) and another set $\gamma^{\prime \nu}$
which  also satisfying Eq. (\ref{Banti}), it is natural to
ask if the two solutions are related. In Minkowski space
Pauli proved that the two sets are always related by a
similarity transformation \cite{Pauli1,Pauli2,Pauli3}.

In curved space-time,  this question is equivalent
to asking what infinitesimal changes $\Delta\gamma^{\nu}$
are possible when $\Delta g^{\nu\kappa}=0$. The
most general solution given in  Eq. (\ref{Btheorem}) is that
$\gamma^{\nu}+\Delta\gamma^{\nu}$ is of the form
$\gamma^{\nu}-[M,\gamma^{\nu}]$  for infinitesimal $M$.
Iterating this shows that the most general solution if of the form
\begin{equation}
\exp(-M)\,\gamma^{\nu}\exp(M).
\end{equation}
Thus any two sets of Dirac matrices
satisfying  anticommutation relations (\ref{Banti}) with
the same metric tensor, must be related by a similarity
transformation. This is used in Sec. II C, where the
spin metric $h$ is the similarity transformation between
$\gamma^{\nu}$ and $\gamma^{\nu\dagger}$ and in Appendix E.

\subsection{The energy-momentum tensor}

 Another application occurs in the computation of  the
energy-momentum tensor in Sec. III B.  There the problem
is to  compute the change in the Dirac
matrices produced by an arbitrary variation in the metric
$\Delta g^{\nu\kappa}$.

\subsection{Proof of the theorem}

To prove Eq. (\ref{Btheorem}) the first step is
to parameterize the most general possible change in the
Dirac matrices as
\begin{eqnarray}
\Delta\gamma^{\nu}
= && 8T^{\nu\alpha}\gamma_{\alpha}
+2 A^{\nu}\gamma_{5}\nonumber\\
+&& B^{\nu\alpha}\gamma_{5}\gamma_{\alpha}
+C^{\nu\alpha\beta}[\gamma_{\alpha},\gamma_{\beta}]
\label{Bexp1}\end{eqnarray}
For a fixed value of $\nu$ the expansion is a linear
combination of 15 traceless matrices.  As already
noted in the above applications, the coefficients
may or may not transform as coordinate tensors depending
upon whether $\Delta g^{\nu\lambda}$ is a tensor.

1. The  coefficient $T^{\nu\alpha}$ can be written
as a trace:
\begin{displaymath}
T^{\nu\kappa}={1\over 32}{\rm Tr}\Big((\Delta
\gamma^{\nu})\gamma^{\kappa}\Big).
\end{displaymath}
The symmetric part of this is
\begin{equation}
T^{\nu\kappa}+T^{\kappa\nu}
={1\over 8}\Delta g^{\nu\kappa}\label{Bsymm}
\end{equation}

2. Next substitute the expansion Eq. (\ref{Bexp1}) into
 Eq. (\ref{BDelta}) to get
\begin{displaymath}
2\Delta g^{\nu\kappa}
=16(T^{\nu\kappa}+T^{\kappa\nu})+
\gamma_{5}X^{\nu\kappa}+Y^{\nu\kappa},
\end{displaymath}
where $X$ and $Y$ denote the matrices
\begin{eqnarray}
X^{\nu\kappa}=&&B^{\nu\alpha}
\big[\gamma_{\alpha},\gamma^{\kappa}\big]
+B^{\kappa\alpha}\big[\gamma_{\alpha},
\gamma^{\kappa}\big]\nonumber\\
Y^{\nu\kappa}=&&C^{\nu\alpha\beta}
\big\{\big[\gamma_{\alpha},\gamma_{\beta}\big],\gamma^{\kappa}\big\}
+C^{\kappa\alpha\beta}
\big\{\big[\gamma_{\alpha},\gamma_{\beta}\big],\gamma^{\nu}\big\}.
\nonumber\end{eqnarray}
Because of Eq. (\ref{Bsymm}) this becomes
\begin{displaymath}
0=\gamma_{5}X^{\nu\kappa}+Y^{\nu\kappa}.
\end{displaymath}
From their definitions,  $\gamma_{5}$ commutes with
$X^{\nu\kappa}$ but anticommutes with $Y^{\nu\kappa}$.
Therefore both matrices vanish
\begin{displaymath}
X^{\nu\kappa}=0\hskip1cm Y^{\nu\kappa}=0.
\end{displaymath}

3. The vanishing of $X$ allows us to extract information
about the coefficients $B^{\nu\alpha}$ by
evaluating the commutator
\begin{displaymath}
0=\big[X_{\mu}^{\nu\kappa},\gamma_{\kappa}\big]
=16B^{\nu\alpha}\gamma_{\alpha}-4\big(g_{\alpha\kappa}
B^{\alpha\kappa}\big)\gamma^{\nu}.
\end{displaymath}
This fixes $B^{\nu\alpha}$ to have the structure
\begin{equation}
B^{\nu\alpha}=-2P\,g^{\nu\alpha},
\label{BP}\end{equation}
where $P$ is unknown.

4. The fact that $Y=0$ will yield a simplification in
the
$C$ coefficients from which it is made.  Define the
anticommutator
\begin{displaymath}
Z^{\kappa}={1\over 4}\big\{Y^{\nu\kappa},
\gamma_{\nu}
\big\}=0.
\end{displaymath}
 Explicit calculation gives
\begin{displaymath}
Z^{\kappa}=3C^{\kappa\alpha\beta}
\big[\gamma_{\alpha},\gamma_{\beta}\big]
+C^{\nu\alpha\beta}\Big(g_{\nu\alpha}
\big[\gamma_{\beta},\gamma^{\kappa}\big]
-g_{\nu\beta}\big[\gamma_{\alpha},\gamma^{\kappa}\big]\Big).
\end{displaymath}
To eliminate the Dirac matrices, compute
\begin{displaymath}
{1\over 16}{\rm
Tr}\big(Z^{\kappa}[\gamma^{\lambda},\gamma^{\rho}
]\big)\!=\!3C^{\kappa[\rho\lambda]}
\!+\!g_{\nu\alpha}\big(C^{\nu[\alpha\rho]}g^{\kappa\lambda}
\!-\!C^{\nu[\alpha\lambda]}g^{\kappa\rho}\big).
\end{displaymath}
Since $Z^{\kappa}=0$, the right hand side must vanish. Therefore the
three-index coefficient has the structure
\begin{equation}
C^{\kappa\rho\lambda}-C^{\kappa\lambda\rho}=
V^{\lambda}\,g^{\kappa\rho}
-V^{\rho}\,g^{\kappa\lambda}.
\label{BV}\end{equation}
with $V^{\lambda}$ unknown.

5. Using the results of Eq. (\ref{Bsymm}), (\ref{BP}),
(\ref{BV}) the expansion in Eq. (\ref{Bexp1})
simplifies to
\begin{eqnarray}
\Delta\gamma^{\nu}
= && {1\over 2}\big(\Delta
g^{\nu\lambda}\big)\gamma_{\lambda}
+4\big(T^{\nu\alpha}-T^{\alpha\nu}\big)\gamma_{\alpha}
+2 A^{\nu}\gamma_{5}\nonumber\\
 -&& 2P\gamma_{5}\gamma^{\nu}
+V^{\beta}[\gamma^{\nu},\gamma_{\beta}]
\label{Bexp2}\end{eqnarray}
Now define a matrix $M$ by
\begin{eqnarray}
M=T^{\alpha\beta}\big[\gamma_{\alpha},\gamma_{\beta}\big]
+A^{\alpha}\gamma_{\alpha}\gamma_{5}+P\gamma_{5}
+V^{\alpha}\gamma_{\alpha}\end{eqnarray}
Then Eq. (\ref{Bexp2}) can be written in the simple form
\begin{equation}
\Delta\gamma^{\nu}
= {1\over 2}\big(\Delta
g^{\nu\lambda}\big)\gamma_{\lambda}
-\big[M,\gamma^{\nu}\big].
\end{equation}
This proves the theorem quoted in Eq. (\ref{Btheorem}).

\section{Contribution of the spin connection to the action}

As mentioned in Sec. II, the full spin connection is parameterized by 16
complex coefficients.  Some parts of the spin connection automatically
cancel out of the fermion action. The spin connection appears in the
action through the combination
\begin{equation}
{i\over 2}\psi^{\dagger}\Big[h\gamma^{\mu}\Gamma_{\mu}-
\Big(h\gamma^{\mu}\Gamma_{\mu}\Big)^{\dagger}\Big]\psi.
\end{equation}
When the general form for the spin connection in Eq. (\ref{spincon}) is
substituted there are a number of cancellations.
To display the result it is convenient to define coefficients which are
traceless on their event indices:
\begin{mathletters}\begin{eqnarray}
\overline{t}_{\mu}^{\alpha\beta}=&& t_{\mu}^{\alpha\beta}
-{1\over 3}\big(\delta_{\mu}^{\alpha}t_{\lambda}^{\lambda\beta}
-\delta_{\mu}^{\beta}t_{\lambda}^{\lambda\alpha}\big)\\
\overline{v}_{\mu}^{\alpha}=&&v_{\mu}^{\alpha}-{1\over 4}
\delta_{\mu}^{\alpha}v_{\lambda}^{\lambda}\\
\overline{a}_{\mu}^{\alpha}=&&a_{\mu}^{\alpha}-{1\over
4}\delta_{\mu}^{\alpha}a_{\lambda}^{\lambda}.
\end{eqnarray}\end{mathletters}
Then the matrix occurring in the action can be written
\begin{equation}
h\gamma^{\mu}\Gamma_{\mu}-
\Big(h\gamma^{\mu}\Gamma_{\mu}\Big)^{\dagger}
=h\big(\gamma^{\mu}A_{\mu}+B\big),
\end{equation}
where the matrix $A_{\mu}$ contains the ``traceless" part of the
coefficients,
\begin{mathletters}\begin{eqnarray}
A_{\mu}=&&2({\rm Re}\,\overline{t}_{\mu}^{\alpha\beta})
\big[\gamma_{\alpha},\gamma_{\beta}\big]
+2({\rm Re}\,\overline{v}_{\mu}^{\alpha})\gamma_{\alpha}\nonumber\\
+&&2i({\rm Im}\,\overline{a}_{\mu}^{\alpha})\gamma_{\alpha}\gamma_{5}
+2i({\rm Im}\,p_{\mu})\gamma_{5}
+2i({\rm Im}\,s_{\mu})I,
\end{eqnarray}
and the matrix $B$ contains the ``traces":
\begin{equation}
B=8i({\rm Im}\,t_{\lambda}^{\lambda\beta})\gamma_{\beta}
+2i({\rm Im}\,v_{\lambda}^{\lambda})I
+2({\rm Re}\,a_{\lambda}^{\lambda})\gamma_{5}.
\end{equation}\end{mathletters}
The full  spin connection in Eq. (\ref{spincon}) contains
16 complex or 32 real parameters for a fixed value of $\mu$. The
combination that occurs in the action contains 16 real
 parameters for a
fixed $\mu$.

\section{Second order Dirac equation and the spin
curvature}

If one iterates the Dirac equation (\ref{Direqn}) the result
 is a second-order
equation
\begin{equation}
0=g^{\mu\nu}\widehat{\nabla}_{\mu}\widehat{\nabla}_{\nu}\psi
+m^{2}\psi+{1\over 2}\big[\gamma^{\mu},\gamma^{\nu}\big]
\nabla_{\mu}\nabla_{\nu}\psi,\label{secDir}
\end{equation}
where the second order covariant derivative is
\begin{displaymath}
\nabla_{\mu}\nabla_{\nu}\psi=\partial_{\mu}(\nabla_{\nu}\psi)
-\Gamma_{\mu\nu}^{\lambda}(\nabla_{\lambda}\psi)
+\Gamma_{\mu}(\nabla_{\nu}\psi).
\end{displaymath}
In the second term of Eq. (\ref{secDir}) the antisymmetric combination of
covariant derivatives   defines  the spin curvature tensor
$\Phi_{\mu\nu}$:
\begin{displaymath}
\nabla_{\mu}\nabla_{\nu}\psi-\nabla_{\nu}\nabla_{\mu}\psi
=\Phi_{\mu\nu}\psi,
\end{displaymath}
where
\begin{equation}
\Phi_{\mu\nu}=\partial_{\mu}\Gamma_{\nu}-\partial_{\nu}\Gamma_{\mu}
+\Gamma_{\mu}\Gamma_{\nu}-\Gamma_{\nu}\Gamma_{\mu}.\end{equation}
Since $\Gamma_{\mu}$ is a linear combination of the Clifford
algebra matrices
$[\gamma_{\alpha},\gamma_{\beta}]$, $\gamma_{\alpha}$,
$\gamma_{\alpha}\gamma_{5}$, $\gamma_{5}$, and $I$ one would expect
that $\Phi_{\mu\nu}$ also contains  these matrices.
(The part of the spin connection that is proportional to the
identity matrix will trivially cancel in $\Phi_{\mu\nu}$ so
that $\Gamma_{\mu}$
and $\widehat{\Gamma}_{\mu}$ produce the same spin curvature tensor.)

The spin curvature tensor can be related to
 the Riemann-Christoffel curvature tensor
\begin{equation}
R_{\mu\nu\kappa}^{\,\cdot\,\cdot\,\cdot\,\lambda}
=\partial_{\mu}\Gamma_{\nu\kappa}^{\lambda}-
\partial_{\nu}\Gamma_{\mu\kappa}^{\lambda}
+\Gamma^{\lambda}_{\mu\alpha}\Gamma^{\alpha}_{\nu\kappa}
-\Gamma_{\nu\alpha}^{\lambda}
\Gamma^{\alpha}_{\mu\kappa},
\end{equation}
in the notation of Schouten
\cite{Schouten}. The commutator of two covariant derivatives
acting on a vector
field is
\begin{displaymath}
\nabla_{\mu}\nabla_{\nu}A^{\lambda}-
\nabla_{\nu}\nabla_{\mu}A^{\lambda}=
R_{\mu\nu\kappa}^{\,\cdot\,\cdot\,\cdot\,\lambda}A^{\kappa}.
\end{displaymath}
 By working out
$\nabla_{\mu}\nabla_{\nu}\gamma^{\lambda}-
\nabla_{\nu}\nabla_{\mu}\gamma^{\lambda}=0$ one obtains the
relation
\begin{equation}0=R_{\mu\nu\kappa}^{\,\cdot\,\cdot\,\cdot\,\lambda}
\gamma^{\kappa}+\big[\Phi_{\mu\nu},\gamma^{\lambda}\big].
\end{equation}
Since $\Phi_{\mu\nu}$ contains no identity component,
this equation requires the spin curvature to be
entirely in the Lorentz subalgebra:
\begin{equation}
\Phi_{\mu\nu}=-{1\over 8}R_{\mu\nu\alpha\beta}
\big[\gamma^{\alpha},\gamma^{\beta}\big].
\end{equation}
When this is substituted into Eq. (\ref{secDir}) the second
 order form of the
Dirac equation becomes
 \begin{equation}
0=g^{\mu\nu}\widehat{\nabla}_{\mu}\widehat{\nabla}_{\nu}\psi
+m^{2}\psi+\big(i\gamma_{5}\widetilde{R}-{1\over 4}IR\big)\psi
\end{equation}
where
\begin{eqnarray}
\tilde{R}=&&{1\over
8\sqrt{-g}}\,\epsilon^{\mu\nu\alpha\beta}R_{\mu\nu\alpha\beta}
\nonumber\\
R=&&g^{\mu\beta}g^{\nu\alpha}R_{\mu\nu\alpha\beta}.\nonumber
\end{eqnarray}

\section{Spin base transformation to vierbein base}

According to the theorem proven in Appendix B, any two sets
of Dirac matrices that solve the anticommutation relations
 must be related by a spin-base similarity
transformation.  Consequently any solution $\gamma^{\mu}$ to
the anticommuation relations is spin-base equivalent to
the vierbein solution:
\begin{equation}
\gamma^{\mu}=Se^{\mu}_{a}\gamma^{a}S^{-1}.\end{equation}
 This Appendix shows how
to construct a  spin base transformation $S$ which does this.

1. Starting with any set $\gamma^{\mu}$, compute
\begin{equation}
\gamma_{5}(x)=-i{\sqrt{-g}\over
4!}\epsilon_{\alpha\beta\mu\nu}
\gamma^{\alpha}\gamma^{\beta}\gamma^{\mu}\gamma^{\nu}.
\end{equation}
Since $[\gamma_{5}(x)]^{2}=I$ the eigenvalues of
$\gamma_{5}(x)$ are $\pm 1$.
It is elementary to find a matrix $S_{1}$ that diagonalizes
$\gamma_{5}(x)$ and therefore transforms it to a constant
matrix. Is important that the transformed matrix be
constant but it need not be diagonal for what follows.
Thus let
\begin{equation}
\gamma_{(5)}=S_{1}^{-1}\gamma_{5}(x)S_{1}
\end{equation}
where $\gamma_{(5)}$ is a constant matrix chosen in some
convenient form. Associated with this constant matrix are
a set of constant Dirac matrices $\gamma^{a}$ which
satisfy  $\{\gamma^{a},\gamma^{b}\}=2\eta^{ab}$
in a representation such that
$i\gamma^{(0)}\gamma^{(1)}
\gamma^{(2)}\gamma^{(3)}=\gamma_{(5)}$.
From $S_{1}$ and the original Dirac matrices construct the
set
\begin{equation}
\gamma^{\prime\mu}(x)=S_{1}^{-1}\gamma^{\mu}(x)S_{1}.
\label{C4}\end{equation}
Each of the new matrices $\gamma^{\prime\mu}(x)$
can be written as a linear combination of the 15 constant
matrices $\gamma^{a}$, $\gamma^{a}\gamma_{(5)}$,
$\gamma_{(5)}$, and $[\gamma^{a},\gamma^{b}]$. However
the vanishing anticommutator
\begin{displaymath}
\big\{\gamma_{(5)},\gamma^{\prime\mu}(x)\big\}=0
\end{displaymath}
means that each $\gamma^{\prime\mu}(x)$ is actually a
linear combination only of the 8 constant  matrices
$\gamma^{a}$ and $\gamma^{a}\gamma_{(5)}$.
Thus write
\begin{equation}
\gamma^{\prime\mu}(x)=V^{\mu}_{a}\gamma^{a}+A^{\mu}_{a}
\gamma^{a}\gamma_{(5)}.\label{C5}
\end{equation}
The anticommutator of these Dirac matrices is
\begin{eqnarray}
\big\{\gamma^{\prime\mu}(x),\gamma^{\prime\nu}(x)\big\}
=&& 2V^{\mu}_{a}V^{\nu}_{b}\eta^{ab}
-2A^{\mu}_{a}A^{\nu}_{b}\eta^{ab}\nonumber\\
+&& \big(V^{\mu}_{a}A^{\nu}_{b}+V^{\nu}_{a}A^{\mu}_{b}\big)
\big[\gamma^{a},\gamma^{b}]\gamma_{(5)}\nonumber
\end{eqnarray}
Since the anticommutator equals $2g^{\mu\nu}$, it follows
that
\begin{eqnarray}
g^{\mu\nu}= && V^{\mu}_{a}V^{\nu}_{b}\eta^{ab}
-A^{\mu}_{a}A^{\nu}_{b}\eta^{ab}\label{C6}\\
0=&&\big(V^{\mu}_{a}A^{\nu}_{b}+V^{\nu}_{a}A^{\mu}_{b}\big)
\big[\gamma^{a},\gamma^{b}]\gamma_{(5)}\nonumber
\end{eqnarray}
The second condition can only be satisfied if
$A^{\mu}_{a}$ is proportional to $V_{a}^{\mu}$.
Therefore set
\begin{displaymath}
A^{\mu}_{a}= V_{a}^{\mu}\tanh\theta\end{displaymath}
where $\theta$ is some function of space-time.
Eq. (\ref{C6}) becomes
\begin{displaymath}
g^{\mu\nu}=V^{\mu}_{a}V^{\nu}_{b}\eta^{ab}(1-\tanh^{2}\theta)
\end{displaymath}
Thus  define the  vierbein
\begin{displaymath}
e^{\mu}_{a}=V^{\mu}_{a}/\cosh\theta,\end{displaymath}
It automatically satisfies
\begin{eqnarray}
g^{\mu\nu}=e^{\mu}_{a}e^{\nu}_{b}\eta^{ab}\end{eqnarray}
The Dirac matrices in Eq. (\ref{C5})  are now
\begin{displaymath}
\gamma^{\prime\mu}=E^{\mu}_{a}
\gamma^{a}(I\cosh\theta+\gamma_{(5)}\sinh\theta)
\end{displaymath}
Since the matrices $\gamma^{(0)}\gamma^{a}$ and
$\gamma^{(0)}\gamma^{a}\gamma_{(5)}$ are self-adjoint, the
matrices $\gamma^{(0)}\gamma^{\prime\mu}$ will only be
self-adjoint if the vierbein $e^{\mu}_{a}$ and the function
$\theta$ are real.

2. To eliminate the $\theta$-dependent chiral rotation, define
another similarity transformation
\begin{displaymath}
S_{2}=I\cosh(\theta/2)-\gamma_{(5)}\sinh(\theta/2).
\end{displaymath}
This matrix has the property
\begin{displaymath}
\gamma^{a}(I\cosh\theta+\gamma_{(5)}\sinh\theta)
=S_{2}\gamma^{a}S_{2}^{-1}.
\end{displaymath}
Therefore
\begin{equation}
\gamma^{\prime\mu}=
S_{2}E^{\mu}_{a}\gamma^{a}S_{2}^{-1}
\label{C8}\end{equation}

3. Combing Eqs. (\ref{C4}) and (\ref{C8}) shows that
the original space-time dependent Dirac matrices can
 be expressed as
\begin{equation}
\gamma^{\mu}=S_{1}S_{2}E^{\mu}_{a}\gamma^{a}S_{2}^{-1}S_{1}^{-1}
\end{equation}
The above procedure gives an explicit method for constructing
the required similarity transformations.

\end{document}